\documentclass[aps,prb,twocolumn,showpacs]{revtex4}
\usepackage{graphics}

\begin{document}

\title{Magneto-Optical Kerr Effect of Iron Thin Films
on Paramagnetic Substrates}

\author{A. Debernardi,$^{a,b}$
I. Galanakis,$^{c}$ M. Alouani,$^{a,}$\footnote{Present Address:
KAVLI Institute for Theoretical Physics, Santa Barbara, Ca
93106-4030, USA} and H. Dreyss\'e~$^a$}

\affiliation{$^a$ Institut de Physique et Chimie des materiaux de
Strasbourg, UMR 7504 du CNRS, Universit\'e Louis Pasteur, 23, rue
du Loess, 67037 Strasbourg, France, European Union \\ $^b$
Istituto Nazionale per la Fisica della Materia (INFM), Udr Trieste
Universit\'a/SISSA, Dip. Fisica Teorica, Strada Costiera 11, 34014
- Trieste, Italy, European Union \\ $^c$ Institut f\"ur
Festk\"orperforschung, Forschungszentrum J\"ulich, D-52425
J\"ulich, Germany,  European Union}

\date{\today}

\begin{abstract}
First principles calculations  of the magnetic properties and
the   magneto-optical Kerr effect (MOKE)
of iron thin films epitaxially grown on the [001] surface of
paramagnetic metals: copper, silver, gold, palladium, and platinum are
presented.
The role of hybridization with the substrate  is investigated
and it is shown  how  the relaxation effects influence the complex Kerr angle.
The results are obtained by means of  the relativistic full-potential
linear muffin-tin method, and
the film is modeled using a slab  geometry within  a supercell technique.
\end{abstract}


\maketitle


The magneto-optical Kerr effect (MOKE)
has become a standard technique for studying the
magnetic properties of a variety of low dimensional systems like
films, surfaces or multilayers~\cite{Ebert}.
This effect occurs when plane polarized light is reflected by a
magnetic material; the light becomes elliptically polarized and the plane of
polarization rotates. Typically, this effect is not very large for the
3$d$ materials but it yields important information for
their magnetic properties as it is
sensitive to the local environment of the 3$d$ atoms.
Although a theoretical investigation has already
been performed to study the effect of the lattice parameter on the
MOKE of bulk 3$d$ metals,\cite{delin} it failed to explain the differences
in MOKE seen for thin films grown on different substrates.
In this contribution we use
the relativistic full-potential linear muffin-tin method\cite{wills}
within the local-spin density approximation\cite{lsda} to
calculate the complex Kerr angle and the magnetic moments
 of an iron monolayer  epitaxially grown on the (001) surface of
five different paramagnetic substrates:
copper, palladium, platinum, gold, and silver.
 These materials, that
belong to two adjacent column of the periodic table,
have a face centered cubic structure  and - due to similar
electronic properties and lattice parameter - allow to
study
the trends of magnetic properties of Fe-layer as a function of
 orbital hybridization and of the lattice constant of the substrate.
We show that the substrate influences the MOKE in a more dramatic way than
the one suggested just by the change of the Fe lattice parameter.


We consider elliptically polarized light and we
define as $\epsilon_K$ the ellipticity
 and as $\theta_K $ the  rotation angle
of the major axis compared to the incident beam.
In the case that the magnetization axis is perpendicular to the surface
the in-plane elements of the conductivity tensor are decoupled from the
perpendicular orientation and
$\sigma_{xx}=\sigma_{yy};\: \sigma_{xy}=-\sigma_{yx}$.
If we consider also the case
of the polar Kerr effect, i.e. the incident
beam is perpendicular to the plane,
then the expression of the complex Kerr angle $\theta_K + i \epsilon_K $
becomes
\begin{equation}
\theta_K +i\epsilon_K =  {-\sigma_{xy}  \over
\sigma_{xx} \sqrt{1+\frac{4\pi i }{\omega}\sigma_{xx} } }
\end{equation}
\noindent
for small Kerr angles as is the case for  the transition metals.
The implementation of the calculation of the conductivity tensor in our
electronic structure method is described in Ref. \onlinecite{delin}.
In our calculations we do not take into account intraband transitions,
which are usually described by the
phenomenological Drude model, but this will affect only the spectra below
0.5 eV and thus will not change the following discussion.

The dielectric tensor $\varepsilon$ is related
to the complex conduction tensor $\sigma\equiv \sigma^{(1)}+i\sigma^{(2)} $
(as usual, $\sigma^{(1)}$ and $\sigma^{(2)}$ denote respectively
 the real part and the coefficient of the
imaginary part)
through the simple relation
\begin{equation}
\varepsilon = 1+{4\pi i \over \omega} \sigma.
\end{equation}
Thus if the complex Kerr angle is know from experiment, one can
extract the dielectric function of a material.


In our calculation  we use a slab structure within a super-cell
geometry. The slab is  composed of
 five layers of paramagnetic material,  with a monolayer of iron
at each extremity of the paramagnetic substrate,
so that no slab-dipole is created. We found that five substrate layers are
sufficient so that
the central layer of Cu (Pd, Pt, Au, or Ag) is bulklike.
In all these calculation we used
the experimental bulk lattice parameters for the substrate
an the plane perpendicular to the grown direction (the values  are displayed in
Tab.~\ref{tavola}).
We also converged the vacuum between two adjacent slabs and
we found that a vacuum of three times the bulk lattice parameter
of the substrate metal is sufficient so that no inter-slab
interactions occur.\cite{us}
Since  the first iron monolayer on the (001) surface
 grows pseudomorphically, we have considered only the
relaxation of the Fe layer
with respect to the rigid substrate.
The equilibrium geometry is obtained by a polynomial fit of
the total energies computed for  different interlayer distances.\cite{noi}
We  found that the relaxation of iron layer is essential
to properly compute magneto-optical effects.\cite{noi}
\begin{table}
\caption{\label{tavola} Calculated spin and orbital magnetic
moments in $\mu_B$ for both the Fe overlayer and the first
noble-metal subsurface layer for all three noble metals. We also
present the experimental lattice bulk parameters $a_{bulk}$ of the
substrate metals and the interlayer distance $d$ between the Fe
layer and the substrate in units of the substrate lattice
constant.}
\begin{ruledtabular}
\begin{tabular}{|l|c|c|c|c|c|c|}
 System & a$_{bulk}$   & $d$
 & m$^{spin}_{Fe}$ &
   m$^{orb}_{Fe}$  & m$^{spin}_{Sub}$ & m$^{orb}_{Sub}$   \\
    &  (\AA) &  (a$_{bulk}$) & ($\mu_B$)&($\mu_B$)&($\mu_B$)&($\mu_B$) \\
\hline Fe/Cu(001) & 3.61 & 0.4703 & 2.728 & .070 & .055 & .008 \\
Fe/Pd(001) & 3.89 & 0.4228 & 3.053 & .090 & .340 & .025 \\
Fe/Pt(001) & 3.92 & 0.4200 & 2.982 & .075 & .308 & .040 \\
Fe/Au(001) & 4.08 & 0.4005 & 2.978 & .070 & .062 & .022 \\
Fe/Ag(001) & 4.09 & 0.4086 & 2.981 & .107 & .013 & .004 \\
\end{tabular}
\end{ruledtabular}
\end{table}
Our results of Fe orbital and spin magnetic moments together with
the first layer substrate induced spin magnetic moments and orbital moments
are reported in Tab.~\ref{tavola}.
For the materials under study
the largest contractions of the Fe layer  along the growth axis
are found for Ag and Au substrates,
to compensate for the larger in-plane expansion of Fe in
the Ag-(Au-)based system compared to the  other ones.
Note that along the [001] direction
the distance between two successive layers for an fcc structure is
0.5 times the bulk lattice parameter;
as the lattice parameter
of the substrate becomes closer to that of the bulk iron the intralayer
distance $d$ approaches the ``ideal" value of 0.5.

\begin{figure}
\scalebox{0.4}{
\includegraphics{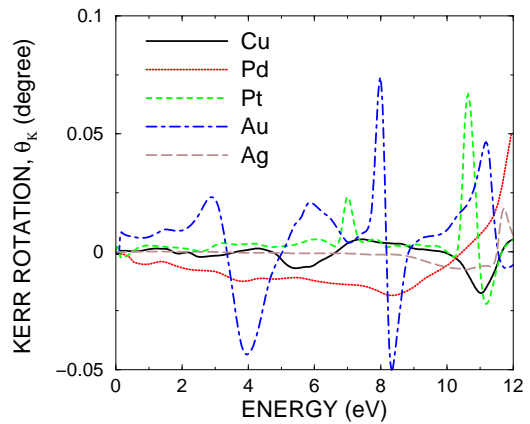}
} \caption{ Magneto-optical Kerr rotation angle  for Fe/Cu(001),
Fe/Pd(001), Fe/Pt(001), Fe/Au(001) and Fe/Ag(001) systems in the
polar configuration. \label{rotation}}
\end{figure}

\begin{figure}
\scalebox{0.4}{
\includegraphics{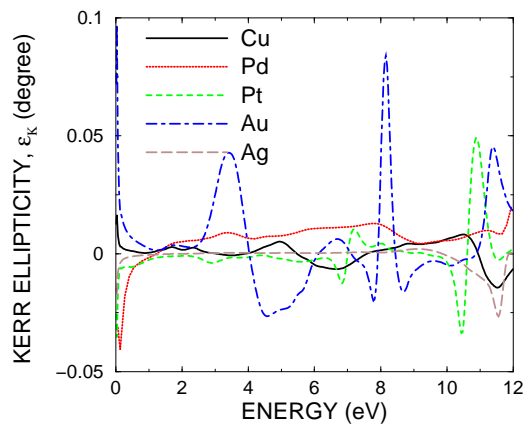}
} \caption{ Magneto-optical Kerr ellipticity  for Fe/Cu(001),
Fe/Pd(001), Fe/Pt(001), Fe/Au(001) and Fe/Ag(001) systems in the
polar configuration. \label{ellipticity}}
\end{figure}

\begin{figure}
\scalebox{0.47}{
\includegraphics{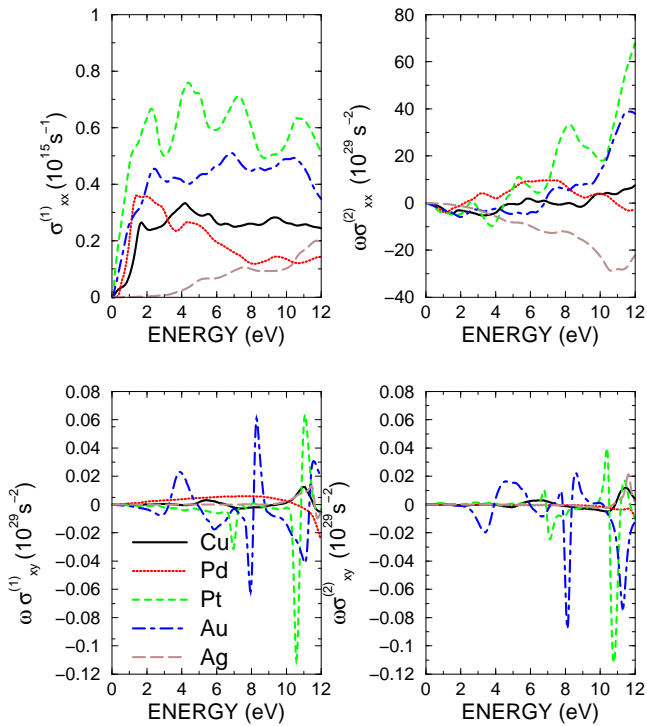}
} \caption{ Calculated optical conductivity for iron monolayer on
different paramagnetic substrate: Cu (solid line), Pd (red line),
Pt (green line), Au (blue line), and Ag (long dashed line). Top
panel: diagonal absorptive optical conductivity (left),
off-diagonal absorptive optical conductivity (right). Bottom
panel:  diagonal dispersive optical conductivity (left),
off-diagonal dispersive optical conductivity (right).
\label{f_cond}}
\end {figure}

The Kerr rotation angle and the ellipticy in the polar configuration
in the case of Fe layer grown on (001) surface of
Cu, Pd, Pt, Au, and Ag~\cite{emrs} substrates
are displayed in
Fig.s~\ref{rotation}, and~\ref{ellipticity}
respectively. For completeness and for further discussion in
Fig.~\ref{f_cond} we
have reported the computed real and imaginary part of in-plane conductivity
tensor.

For Cu substrate
a small enhancement of the Kerr angle $\theta_K$ occurs at $\sim 2.21$ eV
which corresponds to the plasma edge of the bulk copper metal.\cite{kata}
The strong enhancement at an energy of
$\sim 5.5 $ eV and $\sim 6.5 $ eV for $\theta_K$
and $\epsilon_K$, respectively, corresponds to the  depletion
detected already in bulk iron.
Similar feature are also displayed in Fig.~\ref{rotation} for Au substrate
where the signal is magnified according to the fact that in this
system the magnetization of iron and of the top substrate layer
are larger than for Fe/Cu(001) (see Tab.~\ref{tavola}).
Our results for the magneto-optical Kerr effect for Fe/Ag(001) are
presented in Fig.~\ref{rotation}. The richness of structures
displayed in the relaxed spectra of Fe/Cu(001) or Fe/Au(001) is
absent in our computed Kerr  angles for Fe/Ag(001) since, in the
latter system, they became significant only for energies higher
than $\sim 10$ eV. This can be explained by observing the
conductivity tensor of this system is smooth  as a function of
energy. Note that the first Ag-layer display the lower
magnetization among all the substrate under investigation.

In spite that  Fe/Pd(001) and Fe/Pt(001) present the larger magnetization
of the substrate and iron layer,
the Kerr angles present significant variation at energy lower 10 eV only
for Pt system roughly at 6.5 eV due
 to the variation of the off-diagonal
in-plane conductivity in correspondence to this frequency.
The lack of structure in Kerr signal for Pd and Pt system
can be attributed - at least in part -
 to the fact that Pd and Pt elements the valence
$s$-state is empty  (while for the other material is filled by one electron).
These results suggest that the hybridization effects between the
Fe overlayer and the substrate and  the relaxation of the Fe layer
play an equally important role to determine the Kerr effect, and
only their combined effect can explain the observed Kerr spectra.
For these reasons,  a study as the one in Ref.~\onlinecite{delin}
which treats only the change in the lattice parameters is
insufficient to account for the differences in MOKE seen
experimentally for Fe films on different substrates.
To elaborate more on the conclusion of the last paragraph we discuss in detail
our results displayed
in Tab.~\ref{tavola} concerning
 the spin and orbital moments for an Fe monolayer on top of
all five substrates.
We consider firstly the three noble-metal substrates Cu, Au, and Ag,
that belong to the same column of the periodic table.
We remark firstly that the Fe spin moment depends
mainly on the distance between the Fe layer  and
the substrate and thus it is comparable
for the Ag and Au substrates, where also the in-plane Fe expansion
is similar. The Fe orbital moment is comparable for both Cu and Au
substrates, while for the
Ag substrate it is considerably larger. The orbital moment is not
depending only on  the spin moment
of the Fe atom but also on the spin-orbit coupling of the substrate atoms as can
be shown by perturbation theory.\cite{iosif} The most important features
are the
magnetic properties of the first subsurface layer as
they are more sensitive to
hybridization effects than the Fe ones. Firstly we remark that the Fe atoms are
ferromagnetically coupled to the substrate atoms. The Ag spin
moment is considerably
smaller than the Cu and Au ones suggesting that the hybridization between the
Ag 4$d$ and Fe 3$d$ is considerably smaller than between the
Cu 3$d$ (Au 5$d$) and the Fe 3$d$ electrons. This smaller hybridization is
related to the absence of fine structures in the Kerr spectra of Fe/Ag(001) compared
to the other two noble-metal systems. Finally the orbital moment at the Au site is considerably larger
than for the other two noble-metal substrates due to the larger spin-orbit coupling
of the 5$d$ valence electrons compared to the 3$d$ and 4$d$ electrons in the
case of Cu and Ag, respectively.
The trends of the magnetization of  Fe monolayer on Pd and Pt
are opposite to those found for the noble metal since the magnetization
of iron decrease for increasing lattice constant of the substrate.
Pd and Pt have respectively one electron less than
Ag and Au (in 5s or 6s shell). However the fine structure in the Kerr
spectra is larger present in Pt but almost absent in Pd (at least up to 10 eV),
similar trends are found in Ag and Au.
This is a further proof of the role of hybridization of the
$d$-shell as discussed  above for the noble metals.

On summary, we have computed the complex  Kerr angle for a monolayer of iron
pseudomorphically grown on the [001] surface of Cu, Pd, Pt, Au, and Ag.
Our results probe the importance of both the
lattice relaxations and the hybridization effects between the Fe layer and
the substrate on the calculation of the Kerr effect.

\section*{Acknowledgment}

The computer time was granted by IDRIS on the IBM RS6000 (Project No. 001266)
and by Universit\'e Louis Pasteur of Strasbourg on the
SGI Origin-2000 supercomputer.
Financial support for this work has been provided by the TMR-Network of
{\em Interface Magnetism}
(contract ERBFMX96-0089) and the RT Network of {\em Computational
Magnetoelectronics}
(Contract RTN1-1999-00145) of the European Commission.
This research was also supported in part by the National Science
Foundation under Grant No. PHY99-07949.



\end{document}